\documentclass[notitlepage]{report}
\usepackage{titling}
\usepackage[left=1in, right=1in, top=1in, bottom=1in]{geometry}

\usepackage{graphicx}
\usepackage{dcolumn}
\usepackage{bm}
\usepackage{xcolor}
\usepackage{mathtools}
\usepackage{setspace}
\usepackage[colorinlistoftodos]{todonotes}
\usepackage{hyperref}

\title{Multimode optical parametric amplification in the phase-sensitive regime}
\author{Gaetano Frascella$^{1,2}$, Roman V. Zakharov$^{3,4}$, Olga V. Tikhonova$^{3,4}$, \\and Maria V. Chekhova$^{1,2}$}
\date{%
\small    $^1$Max Planck Institute for the Science of Light, Staudtstr. 2, 91058 Erlangen, Germany.\\%
    $^2$University of Erlangen-Nuremberg, Staudtstr. 7/B2, 91058 Erlangen, Germany.\\
    $^3$Physics Department, Moscow State University, Leninskiye Gory 1-2, Moscow 119991, Russia.\\
    $^4$Skobeltsyn Institute of Nuclear Physics, Lomonosov Moscow State University, Moscow 119234, Russia.\\
    \today
}
\begin{document}
\maketitle
\thispagestyle{empty}

\begin{abstract}
Phase-sensitive optical parametric amplification of squeezed states  helps to overcome detection loss and noise and thus increase the robustness of sub-shot-noise sensing. Because such techniques, e.g., imaging and spectroscopy, operate with multimode light, multimode amplification is required. Here we find the optimal methods for multimode phase-sensitive amplification and verify them in an experiment where a pumped second-order nonlinear crystal is seeded with a Gaussian coherent beam. Phase-sensitive amplification is obtained by tightly focusing the seed into the crystal, rather than seeding with close-to-plane waves. This suggests that phase-sensitive amplification of sub-shot-noise images should be performed in the near field. Similar recipe can be formulated for the time and frequency, which makes this work relevant for quantum-enhanced spectroscopy.
\end{abstract}

\vspace{0.5in}

Quantum imaging promises to bring an advantage over classical methods in terms of 
signal-to-noise ratio (SNR)~\cite{Brida:10,Genovese:16,Knyazev:19,Okamoto:20}. Nonclassical light and notably squeezed states of light push the sensitivity to overcome the shot-noise limit in imaging experiments~\cite{Brambilla:08,Samantaray:17}, but their main drawback is fragility. Indeed, squeezing is deteriorated by optical loss and detection inefficiency; therefore, sub-shot-noise experiments require tailored high-transmission equipment and highly efficient detectors. 

Noiseless amplification of squeezed fields prior to optical or detection losses can solve both issues at the same time. This option is realized through phase-sensitive optical parametric amplifiers (OPAs), which consist of the same optical components that generate the squeezed states in the first place. The `noiseless' regime leaves the SNR unchanged from input to output and it is most notoriously achieved with degenerate OPAs. Meanwhile, non-degenerate OPAs amplify without added noise only if both conjugated modes are fed with the input signal. When only one mode is fed with the signal and the other is left with vacuum fluctuations at the input, the added noise translates into a $3$ dB penalty in SNR~\cite{Levenson:93,Ou:93}. Linear amplifiers, based on stimulated emission in a gain medium, suffer from the same problem due to the amplified spontaneous emission.

Amplification without added noise is a longstanding objective in many other fields, such as optical~\cite{Bencheikh:95} and specifically fiber~\cite{Agarwal:14} communications or optomechanics in the microwave range~\cite{Korppi:17}. From seminal papers in the eighties~\cite{Caves:81,Yurke:86}, even quantum metrology has seen rising interest in phase-sensitive OPAs for detection-loss tolerance~\cite{Hudelist:14,Chen:15,Manceau:17PRL,Frascella:19,Frascella:20}. 

The most natural application of multimode OPAs remains imaging; early theoretical~\cite{Kolobov:95} and experimental~\cite{Devaux:95-JOSAB,Choi:99,Devaux:00} works demonstrated the possibility of noiseless amplification of images with bulk crystals. Later, the idea proved efficient also for OPAs based on four-wave mixing~\cite{Corzo:12} and for optical parametric oscillators~\cite{Lopez:08}. In general, phase-sensitive and -insensitive amplifiers have been intensely studied in terms of their transfer function and angular bandwidth~\cite{Gavrielides:87,Choi:99,Corzo:12} and their suppression of twin-beams intensity difference noise~\cite{Marable:98,Liu:19}. To avoid amplification of vacuum noise, the overlap of the OPA eigenmodes with the input radiation is critical~\cite{Lovering:96}, but the requirements to phase-sensitive amplification of images are still missing, just like an understanding of the process in terms of eigenmodes.

In this Letter, we show the conditions for multimode optical parametric amplification in the phase-sensitive regime and confirm them with an experiment. We identify the best strategy to obtain phase-sensitive amplification as projecting a near-field image on the amplifier. Our experiment uses a single mode of an image (a Gaussian coherent seed) fed into a spatially-multimode OPA based on high-gain parametric down-conversion (PDC). The output intensity modulation quantifies the sensitivity to the input phase and we study its dependence on the initial beam divergence and tilt with respect to the pump direction. We also discuss the time-domain analogy of this experiment. 

A spatially multimode OPAs is described by the Hamiltonian

\begin{equation}
\hat{{\cal H}}=i\hbar\Gamma\iint d\boldsymbol{x}_{s}d\boldsymbol{x}_{i}F\left(\boldsymbol{x}_{s},\boldsymbol{x}_{i}\right)a^{\dagger}\left(\boldsymbol{x}_{s}\right)a^{\dagger}\left(\boldsymbol{x}_{i}\right)+{\rm h.\,c.\,,}\label{eq:multimode_Hamiltonian}
\end{equation}
where $\Gamma$ is a coupling parameter proportional to the pump field amplitude, the length of the medium and the nonlinearity, while $\boldsymbol{x}$ can be either a transverse spatial coordinate $\boldsymbol{\rho}$ (near-field) or the Fourier-related transverse wave-vector $\boldsymbol{q}$ (far-field). To quantify the amount of amplification, one defines the parametric gain of the OPA as $G=\intop\Gamma\,dt$.

The two-photon amplitude (TPA) $F\left(\boldsymbol{x}_{s},\boldsymbol{x}_{i}\right)$ describes the joint probability amplitude of emitting photons into the signal mode with coordinate $\boldsymbol{x}_{s}$ and the idler mode with coordinate $\boldsymbol{x}_{i}$. The near-field TPA $F_{{\rm near}}\left(\boldsymbol{\rho}_{s},\boldsymbol{\rho}_{i}\right)$ and the far-field one $F_{{\rm far}}\left(\boldsymbol{q}_{s},\boldsymbol{q}_{i}\right)$ are related by a Fourier transform. Eq.(~\ref{eq:multimode_Hamiltonian}) is valid also for multiple modes in time or frequency domain, if $\boldsymbol{x}$ is time or frequency.

A simple case to understand our viewpoint on phase-sensitive amplification is the one of very short medium and very large pump beam waist. Then, the far-field TPA scales as a delta function $F_{{\rm far}}\sim\delta\left(\boldsymbol{q}_{s}+\boldsymbol{q}_{i}\right)$, and the Fourier transform yields the near-field TPA in the form $F_{{\rm near}}\sim\delta\left(\boldsymbol{\rho}_{s}-\boldsymbol{\rho}_{i}\right)$.
The Hamiltonian in Eq.(~\ref{eq:multimode_Hamiltonian}) is simplified to

\begin{equation}
\hat{{\cal H}}_{{\rm far}}=i\hbar\Gamma\int d\boldsymbol{q}_{s}a^{\dagger}\left(\boldsymbol{q}_{s}\right)a^{\dagger}\left(-\boldsymbol{q}_{s}\right)+{\rm h.\,c.\,,}\\
\label{eq:delta_Hamiltonian_far}
\end{equation}
\begin{equation}
\hat{{\cal H}}_{{\rm near}}=i\hbar\Gamma\int d\boldsymbol{\rho}_{s}\left[a^{\dagger}\left(\boldsymbol{\rho}_{s}\right)\right]^{2}+{\rm h.\,c.\,.}
\label{eq:delta_Hamiltonian_near}
\end{equation}
 In the far field (Eq.(~\ref{eq:delta_Hamiltonian_far})), an analogy can be made with the two-mode-OPA Hamiltonian, while in the near field (Eq.~(\ref{eq:delta_Hamiltonian_near})) with the single-mode-OPA one. 
 
 \begin{figure}[b]
\centering
\includegraphics[width=0.5\paperwidth]{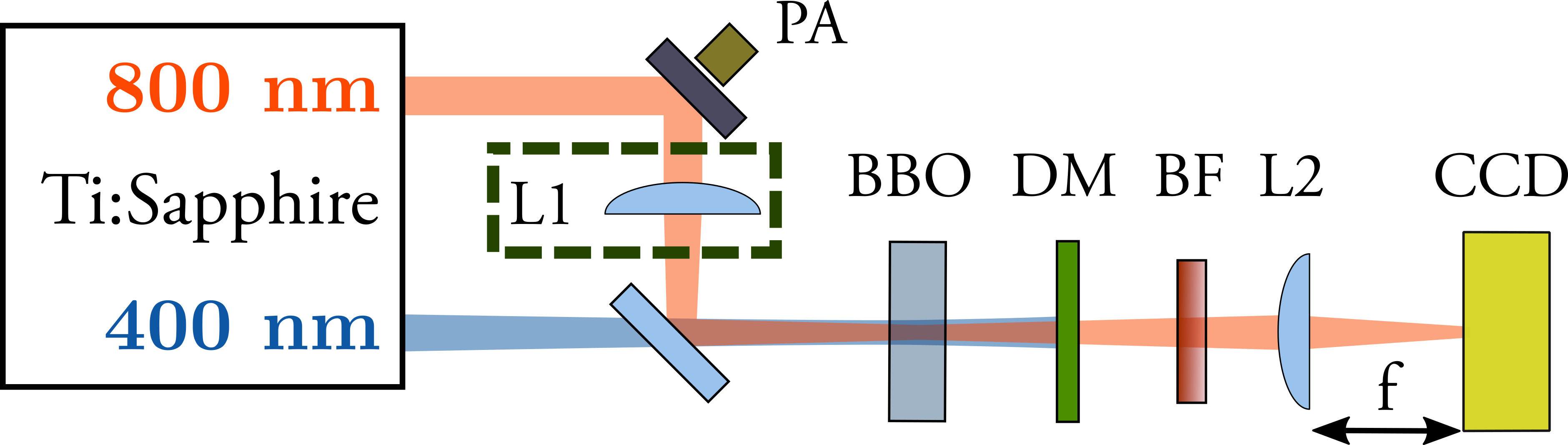}
\caption{Experimental setup for optical parametric amplification of a large-divergence (with the lens in the dashed box) or a small-divergence  (without it) seed. PA, piezoelectric actuator; L1,L2 lenses; BBO, $\beta$-barium borate crystal; DM, dichroic mirror; BF, bandpass filter; CCD, charge-coupled device camera. \label{fig:setup} }
\end{figure}

 For a two-mode-OPA, if one input is fed with a coherent beam and the other with vacuum, the coherent beam is amplified and a conjugated beam appears regardless of the input phase (phase-insensitive or phase-preserving~\cite{Clerk:10}). If both signal and idler inputs are fed with coherent states with two independent phases, the output beams will show amplification or de-amplification depending of the sum of the phases of the input beams (phase-sensitive) and the original phases are not preserved at the output. This means that, in the far field, the OPA described with Hamiltonian (\ref{eq:delta_Hamiltonian_far}) is phase-sensitive if plane wave modes at conjugated transverse wavevectors at $\boldsymbol{q}_{s}$ and $\boldsymbol{q}_{i}=-\boldsymbol{q}_{s}$ are fed with coherent beams simultaneously.
In the single-mode case, the amplification is shown to be always dependent on the phase of the input coherent state. Therefore, seeding in the near field, i.e. at one point of the OPA, results in phase-sensitive amplification.

We realize experimentally far-field and near-field seeding of a multimode OPA, respectively, with a small- and large-divergence coherent beam and we test the modulation of the signal output intensity with the phase. We also vary the seed central angle to show the transition between seeding conjugated and non-conjugated input transverse wavevectors. In our experiment, the OPA is a $\chi^{(2)}$-nonlinear crystal working in the collinear degenerate regime. Signal and idler beams are discerned by considering the emission in one half of the spatial emission spectrum as the signal and the other half as the idler, given the anti-correlation of the two beams in transverse wavevector.

Fig.~\ref{fig:setup} shows the experimental setup. The coherent beam comes from a Ti:Sapphire laser at 800 nm producing $1.5$ ps pulses at $5$ kHz repetition rate. The phase of this beam is scanned linearly in time (triangle wave) with the piezoelectric actuator PA, while the divergence is changed by inserting or removing lens L1. The intensity full width at half maximum (FWHM) is $80\pm10$ $\mu$m with lens L1 and $2.3\pm0.1$ mm without the lens, resulting in a divergence of $3.8\pm0.5$ mrad and $0.13\pm0.01$ mrad, respectively. 

The pump is the second harmonic of the same laser and its waist position is the same as the one of the coherent seed. Here, we place the $2$-mm $\beta$-barium borate crystal BBO. The pump has intensity FWHM $240\pm10$~$\mu$m and an average power of $65$ mW, and it is rejected with the dichroic mirror DM. We obtain a value of $G=3.2\pm0.3$ for the OPA parametric gain, measured from the nonlinear dependence of the PDC intensity on the pump power~\cite{Spasibko:12}. We filter the output radiation using a bandpass filter with central wavelength $800$ nm and bandwidth $10$ nm to reduce the contribution of the non-seeded PDC. In the case of a large-divergence seed, we record the spatial spectrum of the crystal emission as a function of time by placing the charge-coupled device (CCD) camera in the Fourier plane of lens L2. To avoid saturation in the case of a small divergence (sharp wavevectors spectrum), we move the CCD camera slightly away from the Fourier plane. 
\begin{figure}[h]
\centering
\includegraphics[width=0.5\paperwidth]{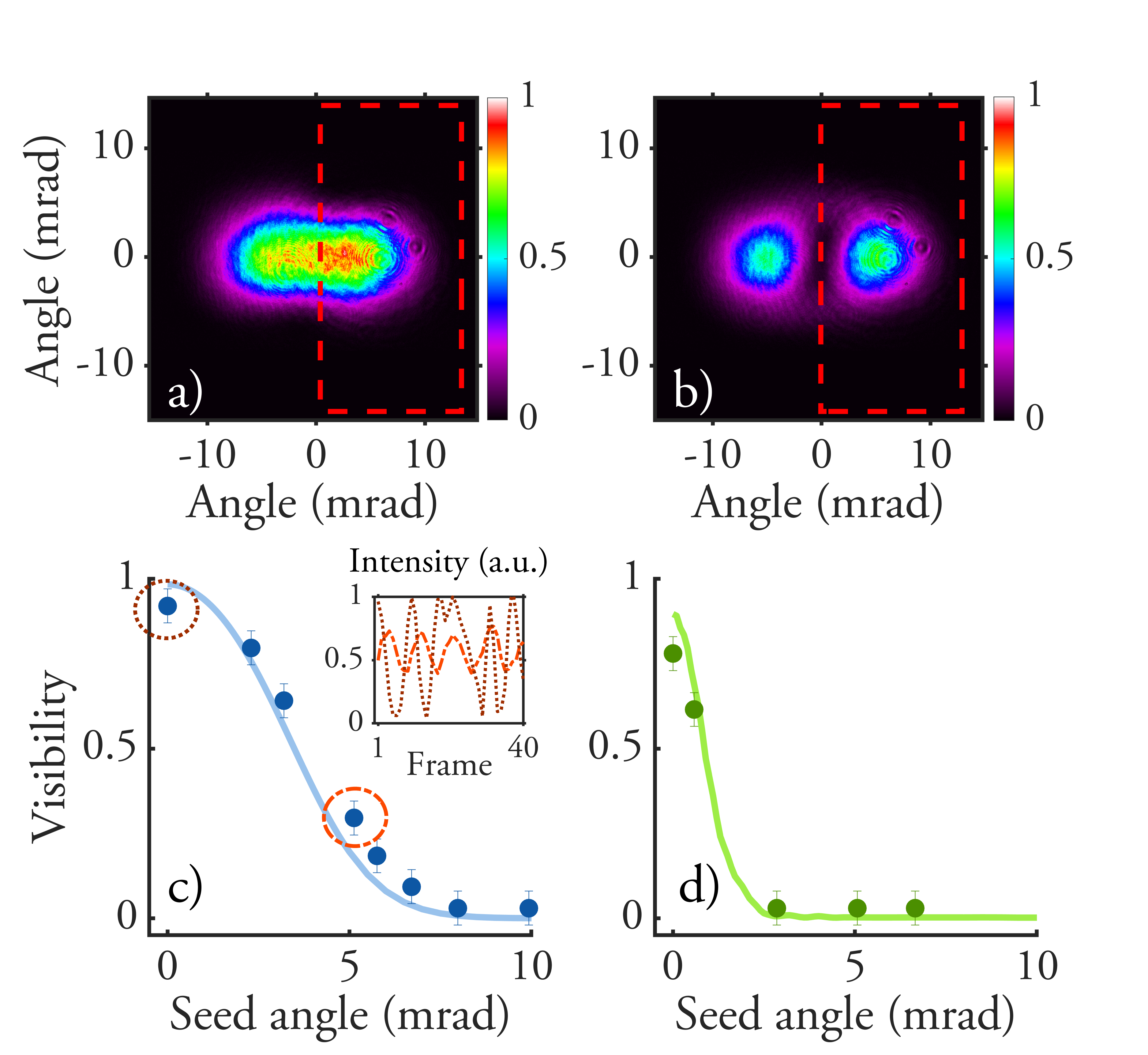}\caption{Two-dimensional angular spectrum at the OPA output with amplified (a) and de-amplified coherent beam (b) with large divergence and central angle 5 mrad. The spectra are integrated over the dashed red region to obtain the signal intensity. Visibility of the signal intensity phase-sensitive modulation (the inset showing the time dependence for two marked points) versus the seed central angle for large (c) and small (d) divergence.\label{fig:exp-results}}
\end{figure}

The upper panels of Fig.~\ref{fig:exp-results} show two spatial spectra of the OPA output in the case of near-field seeding with a central angle of $5$ mrad: the coherent beam is amplified at some phase (a), while it is de-amplified if its phase is shifted by $\pi/2$ (b). The spectrum is integrated over the red dashed rectangle to obtain the signal intensity and its variation with the phase. Just like the interferometric visibility is a practical way to measure first-order coherence, the phase-dependent modulation of the signal photon number helps distinguishing between the phase-sensitive and -insensitive regime of parametric amplification.

The lower panels of Fig.~\ref{fig:exp-results} show the results of this procedure, carried out for several central angles of the seed and for both near- (c) and far-field seeding (d). 
The phase-sensitive modulation drops for large seed central angles with a FWHM of $\sim4$ mrad in the large-divergence case. Meanwhile, the FWHM is $\sim1$ mrad in the small-divergence case. The experiment (points in Fig.~\ref{fig:exp-results} c-d) agrees well with the simulation (lines in Fig.~\ref{fig:exp-results} c-d), which we implement for the general case of a non-delta TPA. 

Indeed, these results can be better understood in terms of the eigenmodes of the OPA: the coherent beam is well amplified in the phase-sensitive regime if its divergence allows to populate modes that are coupled in transverse wavevectors. The eigenmodes are obtained from the Schmidt decomposition of the TPA, whose symmetry in the azimuthal angle (valid for instance for PDC in nonlinear crystals or four-wave mixing in Rb vapor cells) allows to write, both for the far~\cite{Sharapova:15} and near~\cite{Zakharov:18} fields,

\begin{equation}
F\left(\boldsymbol{x}_{s},\boldsymbol{x}_{i}\right)=\sum_{n,p}\sqrt{\lambda_{n,p}}U_{n,p}\left(\boldsymbol{x}_{s}\right)V_{n,p}\left(\boldsymbol{x}_{i}\right),\label{eq:Schmidt_decomp}
\end{equation}
where $n$ and $p$ are, respectively, azimuthal and radial indices, $\lambda_{n,p}$ are the Schmidt weights and $U$, $V$ are, respectively, Schmidt signal and idler modes. This is a simplified model, in which the shapes of the modes do not depend on the gain; a rigorous theory, proved by experiment, shows that the mode shapes change very little~\cite{Sharapova:20}. 

Given this decomposition, the Hamiltonian can be rewritten as a weighted sum of the products of collective signal and idler creation operators. Each term of the sum corresponds to a two-mode OPA with the parametric gain $G_{n,p}\equiv G\sqrt{\lambda_{n,p}}$.  If the parametric gain is increased, the effective number of modes is reduced~\cite{Sharapova:15}, and so will be the imaging resolution. The choice of these two parameters is always a compromise.

This multimode formalism is powerful because it allows to calculate the moments of the observables of interest, and, in particular, the output signal photon number $\hat{N_{s}}$ averaged for an input state of the OPA in a specific mode~\cite{Fabre:20}.

For a strong coherent input beam with complex amplitude $\alpha$ in the spatial mode $f\left(\boldsymbol{x}\right)$, we calculate the average signal photon number:

\begin{equation}
\begin{array}{c}
\left\langle \alpha,f\right|\hat{N}_{s}\left|\alpha,f\right\rangle \approx\sum_{n,p}\left|\alpha\right|^{2}\left[\left|\beta_{n,p}\right|^{2}\cosh^{2}G_{n,p}\right.\\
+\left|\beta'_{n,p}\right|^{2}\sinh^{2}G_{n,p}+\sinh2G_{n,p}\left|\beta_{n,p}\right|\left|\beta'_{n,p}\right|\\
\left.\times\cos\left(2\arg\alpha-\arg\beta{}_{n,p}-\arg\beta'_{n,p}\right)\right],
\end{array}\label{eq:avg-signal}
\end{equation}
where $\beta_{n,p}=\intop d\boldsymbol{x}f^{*}\left(\boldsymbol{x}\right)U_{n,p}\left(\boldsymbol{x}\right)$ and $\beta'_{n,p}=\intop d\boldsymbol{x}f^{*}\left(\boldsymbol{x}\right)V_{n,p}\left(\boldsymbol{x}\right)$ are the overlap integrals of the seed with signal/idler modes. The non-seeded PDC emission is neglected, since its number of photons does not depend on $\alpha$.  

The phase-insensitive part of Eq.~(\ref{eq:avg-signal}) is described by the first two terms in the square brackets. If the coherent input mode overlaps with the signal (idler) mode $\left(n,p\right)$, $\left|\alpha\right|^{2}\sinh^{2}G_{n,p}$ photons will be added to the mode, previously populated by $\left|\alpha\right|^{2}$ (zero) photons. Meanwhile, the last and phase-sensitive term in Eq.(~\ref{eq:avg-signal}) depends on the phase $\arg\alpha$ of the coherent beam only if, for at least one value of $n$ and $p$, the overlap integrals $\beta{}_{n,p}$ and $\beta'_{n,p}$ have simultaneously non-zero modulus. In other words, the condition of phase-sensitive amplification is valid only if the coherent beam mode overlaps with signal and idler modes simultaneously. 

This overlap condition is easy to satisfy in the near field, where the signal and idler modes are non-zero at the same points; it is more difficult in the far field, where these modes peak at opposite wavevectors~\cite{Liu:19,Choi:99}. This suggests that one should be imaging the seed beam waist in the photon creation region of the OPA to obtain noiseless amplification, or, ultimately, one should seed by a spherical wave rather than by a plane wave.

Two other important aspects of Eq.(~\ref{eq:avg-signal}) are: i) for a fixed $n$ and $p$, each term in the squared brackets is proportional to an exponential function of $G_{n,p}$, which means that modes with higher weights are stronger amplified~\cite{Huo:20}; ii) for a degenerate OPA, this formula can be simplified because the signal and idler modes coincide in modulus $\left|U_{n,p}\right|=\left|V_{n,p}\right|$ and the relation between the overlap integrals $\beta_{n,p}^{'}=\beta_{-n,p}$ can be derived.

\begin{figure}
\centering
\includegraphics[height=0.17\paperheight]{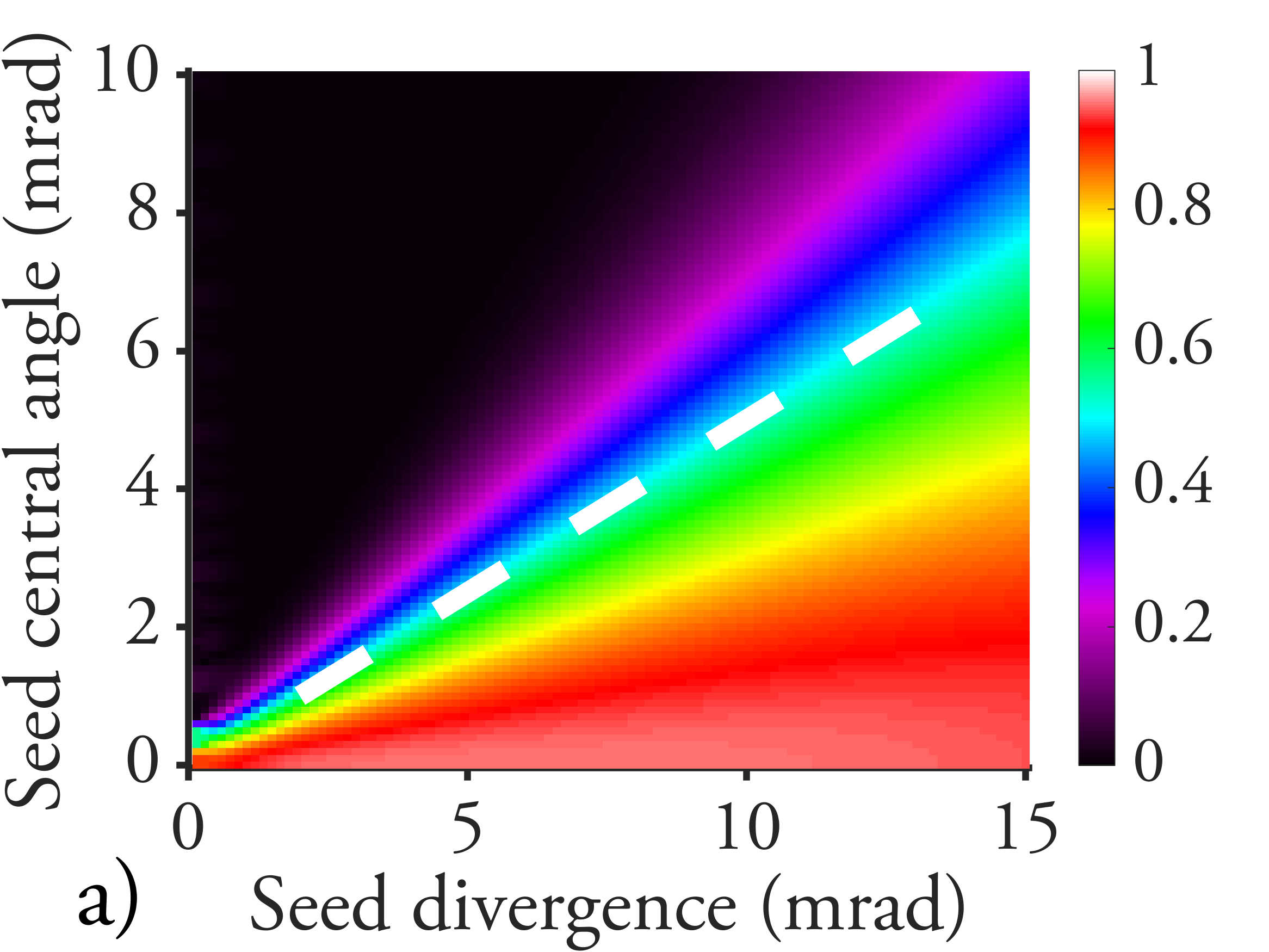}~\includegraphics[height=0.17\paperheight]{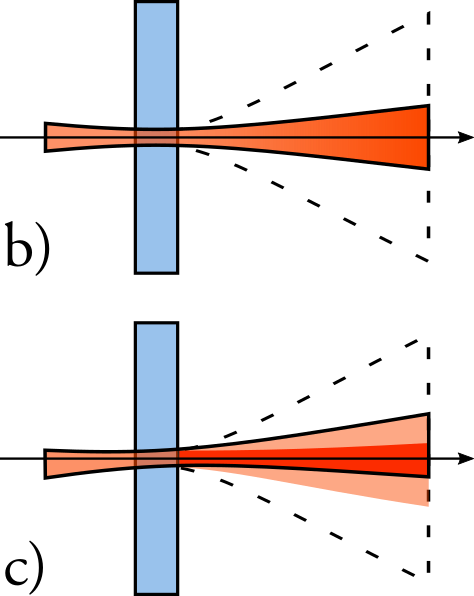}\caption{a) Simulated phase-sensitive modulation visibility as a function of the seed divergence and central angle with respect to the pump direction; along the white dashed line the central angle is half the divergence. b) The seed (solid line) along the pump direction (arrow) always undergoes phase-sensitive amplification, with a divergence limited by the OPA non-seeded emission (dashed line). c) When the seed is tilted, the amplification is partly phase-sensitive (stronger shading) and partly phase-insensitive (lighter shading). \label{fig:vis-divergence}}
\end{figure}

Fig.~\ref{fig:vis-divergence} a shows the calculated visibility of phase-sensitive intensity modulation for a Gaussian seed as a function of divergence and central angle for type-I degenerate PDC with the same parameters as in our experiment. The seed spatial spectrum has the form $f\left(\boldsymbol{q}\right)\propto\exp\left[-\lambda^{2}\left|\boldsymbol{q}-\boldsymbol{q}_{0}\right|^{2}/\left(2\pi^{2}\theta^{2}\right)\right]$ with divergence $\theta$ and transverse wavevector $\boldsymbol{q}_{0}$.

For a seed beam propagating along the pump direction, the amplification depends strongly on the coherent seed phase for any value of the seed divergence, but a maximum is obtained when the seed matches the full angular range of the non-seeded emission of the OPA. Fig.~\ref{fig:vis-divergence} b shows a sketch of seed (solid line) propagating along the pump direction (arrow); the seed is amplified in the phase-sensitive regime even if it does not match the angular bandwidth of the non-seeded PDC emission (dashed line) from the nonlinear crystal.

For a seed tilted with respect to the pump direction, the sensitivity to the seed phase is high when the central angle is smaller than half the divergence (below the white dashed line in Fig.~\ref{fig:vis-divergence} a). Fig.~\ref{fig:vis-divergence} c shows a sketch of this case: if the seed can populate modes with opposite wavevectors, there will be a part of phase-sensitive amplification (stronger shading) with good intensity modulation but also a phase-insensitive part (lighter shading).

The results described here for OPAs with multiple modes in space have a time-domain analogy. Indeed, photons of a pair are correlated both in position (near-field coordinate) and in time; meanwhile, they are anti-correlated in momentum (far-field coordinate) and frequency. Therefore, to have phase-sensitive (noiseless) amplification one has to amplify a time pattern, rather than a spectrum. A straightforward conclusion is that short pulses will be always amplified noiselessly, but with the limitation imposed by the frequency bandwidth of the OPA. This conclusion is very useful for enhanced sub-shot-noise spectroscopy, because complex temporal profiles can be studied through the analysis of the phase-sensitive intensity modulation. 

Our work is readily applicable in the field of quantum imaging, where phase-sensitive amplification before detection can provide loss tolerance~\cite{Knyazev:19}. For a single mode of an image, both experimental and theoretical evidence point towards near-field or high-divergence seeding to obtain phase-sensitive amplification. In contrast to far-field amplification, projecting the near-field image on the OPA gives phase-sensitive intensity modulation tolerant to the input central angle. Our calculations can be adapted to real-world case scenarios and help to model theoretically experiments conducted with multimode OPAs.

Our experiment is based on a bulk-crystal OPA, but these results on multimode phase-sensitive amplification are valid also for systems based on four-wave mixing. The main difficulty for the latter is that amplified signal is hard to be distinguished from the pump; therefore, two far-field replicas of the image are fed into conjugate modes with wavevectors slightly off from the pump direction. Meanwhile, nonlinear crystals can be used in the collinear regime and dichroic mirrors easily reject the pump radiation.

The above analysis of the contribution of different modes to the phase-sensitive modulation can prove very useful in the context where multimode OPAs are used to achieve sub-shot-noise precision to phase shifts. This is the case of SU(1,1) interferometers in the seeded regime~\cite{Plick:10,Manceau:17} and the squeezing-assisted interferometer with output optical parametric amplification~\cite{Frascella:20}.

The advantage of squeezing, especially in quantum imaging and sensing, is mostly limited by detection efficiency. Our work makes a step further in the analysis of multimode phase-sensitive parametric optical amplification, which, if employed before detection, can compensate for inefficiencies.

\section*{Acknowledgments}
This research has been supported by the Interdisciplinary Scientific and Educational School of Lomonosov Moscow State University ``Photonic and Quantum technologies. Digital medicine".
\bibliographystyle{plain}
\bibliography{nearfieldPSA}

\begin{thebibliography}{10}

\bibitem{Agarwal:14}
Anjali Agarwal, James~M. Dailey, Paul Toliver, and Nicholas~A. Peters.
\newblock Entangled-pair transmission improvement using distributed
  phase-sensitive amplification.
\newblock {\em Phys. Rev. X}, 4:041038, Dec 2014.

\bibitem{Bencheikh:95}
K.~Bencheikh, O.~Lopez, I.~Abram, and J.~A. Levenson.
\newblock Improvement of photodetection quantum efficiency by noiseless optical
  preamplification.
\newblock {\em Applied Physics Letters}, 66(4):399--401, 1995.

\bibitem{Brambilla:08}
E.~Brambilla, L.~Caspani, O.~Jedrkiewicz, L.~A. Lugiato, and A.~Gatti.
\newblock High-sensitivity imaging with multi-mode twin beams.
\newblock {\em Phys. Rev. A}, 77:053807, May 2008.

\bibitem{Brida:10}
G.~Brida, M.~Genovese, and I.~Ruo-Berchera.
\newblock Experimental realization of sub-shot-noise quantum imaging.
\newblock {\em Nat Photon}, 4(4):227--230, 04 2010.

\bibitem{Caves:81}
Carlton~M. Caves.
\newblock Quantum-mechanical noise in an interferometer.
\newblock {\em Phys. Rev. D}, 23:1693--1708, Apr 1981.

\bibitem{Chen:15}
Bing Chen, Cheng Qiu, Shuying Chen, Jinxian Guo, L.~Q. Chen, Z.~Y. Ou, and
  Weiping Zhang.
\newblock Atom-light hybrid interferometer.
\newblock {\em Phys. Rev. Lett.}, 115:043602, Jul 2015.

\bibitem{Choi:99}
Sang-Kyung Choi, Michael Vasilyev, and Prem Kumar.
\newblock Noiseless optical amplification of images.
\newblock {\em Phys. Rev. Lett.}, 83:1938--1941, Sep 1999.

\bibitem{Clerk:10}
A.~A. Clerk, M.~H. Devoret, S.~M. Girvin, Florian Marquardt, and R.~J.
  Schoelkopf.
\newblock Introduction to quantum noise, measurement, and amplification.
\newblock {\em Rev. Mod. Phys.}, 82:1155--1208, Apr 2010.

\bibitem{Corzo:12}
N.~V. Corzo, A.~M. Marino, K.~M. Jones, and P.~D. Lett.
\newblock Noiseless optical amplifier operating on hundreds of spatial modes.
\newblock {\em Phys. Rev. Lett.}, 109:043602, Jul 2012.

\bibitem{Devaux:95-JOSAB}
F.~Devaux and E.~Lantz.
\newblock Parametric amplification of a polychromatic image.
\newblock {\em J. Opt. Soc. Am. B}, 12(11):2245--2252, Nov 1995.

\bibitem{Devaux:00}
Fabrice Devaux and Eric Lantz.
\newblock Gain in phase sensitive parametric image amplification.
\newblock {\em Phys. Rev. Lett.}, 85:2308--2311, Sep 2000.

\bibitem{Fabre:20}
C.~Fabre and N.~Treps.
\newblock Modes and states in quantum optics.
\newblock {\em Rev. Mod. Phys.}, 92:035005, Sep 2020.

\bibitem{Frascella:19}
G.~Frascella, E.~E. Mikhailov, N.~Takanashi, R.~V. Zakharov, O.~V. Tikhonova,
  and M.~V. Chekhova.
\newblock Wide-field {SU(1,1)} interferometer.
\newblock {\em Optica}, 6(9):1233--1236, Sep 2019.

\bibitem{Frascella:20}
Gaetano {Frascella}, Sascha {Agne}, Farid~Ya. {Khalili}, and Maria~V.
  {Chekhova}.
\newblock {Overcoming detection inefficiency in squeezing-assisted
  interferometers}.
\newblock {\em arXiv e-prints}, page arXiv:2005.08843, May 2020.

\bibitem{Gavrielides:87}
A.~Gavrielides, P.~Peterson, and D.~Cardimona.
\newblock Diffractive imaging in three‐wave interactions.
\newblock {\em Journal of Applied Physics}, 62(7):2640--2645, 2020/10/04 1987.

\bibitem{Genovese:16}
Marco Genovese.
\newblock Real applications of quantum imaging.
\newblock {\em Journal of Optics}, 18(7):073002, jun 2016.

\bibitem{Hudelist:14}
F.~Hudelist, Jia Kong, Cunjin Liu, Jietai Jing, Z.~Y. Ou, and Weiping Zhang.
\newblock Quantum metrology with parametric amplifier-based photon correlation
  interferometers.
\newblock {\em Nature Communications}, 5:3049 EP --, 01 2014.

\bibitem{Huo:20}
Nan Huo, Yuhong Liu, Jiamin Li, Liang Cui, Xin Chen, Rithwik Palivela, Tianqi
  Xie, Xiaoying Li, and Z.~Y. Ou.
\newblock Direct temporal mode measurement for the characterization of
  temporally multiplexed high dimensional quantum entanglement in continuous
  variables.
\newblock {\em Phys. Rev. Lett.}, 124:213603, May 2020.

\bibitem{Knyazev:19}
Eugene Knyazev, Farid~Ya. Khalili, and Maria~V. Chekhova.
\newblock Overcoming inefficient detection in sub-shot-noise absorption
  measurement and imaging.
\newblock {\em Opt. Express}, 27(6):7868--7885, Mar 2019.

\bibitem{Kolobov:95}
Mikhail~I. Kolobov and Luigi~A. Lugiato.
\newblock Noiseless amplification of optical images.
\newblock {\em Phys. Rev. A}, 52:4930--4940, Dec 1995.

\bibitem{Levenson:93}
J.~A. Levenson, I.~Abram, Th. Rivera, and Ph. Grangier.
\newblock Reduction of quantum noise in optical parametric amplification.
\newblock {\em J. Opt. Soc. Am. B}, 10(11):2233--2238, Nov 1993.

\bibitem{Liu:19}
Shengshuai Liu, Yanbo Lou, and Jietai Jing.
\newblock Interference-induced quantum squeezing enhancement in a two-beam
  phase-sensitive amplifier.
\newblock {\em Phys. Rev. Lett.}, 123:113602, Sep 2019.

\bibitem{Lopez:08}
L.~Lopez, N.~Treps, B.~Chalopin, C.~Fabre, and A.~Ma\^{\i}tre.
\newblock Quantum processing of images by continuous wave optical parametric
  amplification.
\newblock {\em Phys. Rev. Lett.}, 100:013604, Jan 2008.

\bibitem{Lovering:96}
D.~J. Lovering, J.~A. Levenson, P.~Vidakovic, J.~Webj\"{o}rn, and P.~St.~J.
  Russell.
\newblock Noiseless optical amplification in quasi-phase-matched bulk lithium
  niobate.
\newblock {\em Opt. Lett.}, 21(18):1439--1441, Sep 1996.

\bibitem{Manceau:17}
Mathieu Manceau, Farid Khalili, and Maria Chekhova.
\newblock Improving the phase super-sensitivity of squeezing-assisted
  interferometers by squeeze factor unbalancing.
\newblock {\em New Journal of Physics}, 19(1):013014, 2017.

\bibitem{Manceau:17PRL}
Mathieu Manceau, Gerd Leuchs, Farid Khalili, and Maria Chekhova.
\newblock Detection loss tolerant supersensitive phase measurement with an
  $\rm{SU(1,1)}$ interferometer.
\newblock {\em Phys. Rev. Lett.}, 119:223604, Nov 2017.

\bibitem{Marable:98}
Michael~L. Marable, Sang-Kyung Choi, and Prem Kumar.
\newblock Measurement of quantum-noise correlations in parametric image
  amplification.
\newblock {\em Opt. Express}, 2(3):84--92, Feb 1998.

\bibitem{Korppi:17}
C.~F. Ockeloen-Korppi, E.~Damsk\"agg, J.-M. Pirkkalainen, T.~T. Heikkil\"a,
  F.~Massel, and M.~A. Sillanp\"a\"a.
\newblock Noiseless quantum measurement and squeezing of microwave fields
  utilizing mechanical vibrations.
\newblock {\em Phys. Rev. Lett.}, 118:103601, Mar 2017.

\bibitem{Okamoto:20}
Ryo Okamoto, Yuta Tokami, and Shigeki Takeuchi.
\newblock Loss tolerant quantum absorption measurement.
\newblock {\em New Journal of Physics}, 2020.

\bibitem{Ou:93}
Z.~Y. Ou, S.~F. Pereira, and H.~J. Kimble.
\newblock Quantum noise reduction in optical amplification.
\newblock {\em Phys. Rev. Lett.}, 70:3239--3242, May 1993.

\bibitem{Plick:10}
William~N Plick, Jonathan~P Dowling, and Girish~S Agarwal.
\newblock Coherent-light-boosted, sub-shot noise, quantum interferometry.
\newblock {\em New Journal of Physics}, 12(8):083014, 2010.

\bibitem{Samantaray:17}
Nigam Samantaray, Ivano Ruo-Berchera, Alice Meda, and Marco Genovese.
\newblock Realization of the first sub-shot-noise wide field microscope.
\newblock {\em Light: Science \& Applications}, 6(7):e17005--e17005, 2017.

\bibitem{Sharapova:15}
P.~Sharapova, A.~M. P\'erez, O.~V. Tikhonova, and M.~V. Chekhova.
\newblock Schmidt modes in the angular spectrum of bright squeezed vacuum.
\newblock {\em Phys. Rev. A}, 91:043816, Apr 2015.

\bibitem{Sharapova:20}
P.~R. Sharapova, G.~Frascella, M.~Riabinin, A.~M. P\'erez, O.~V. Tikhonova,
  S.~Lemieux, R.~W. Boyd, G.~Leuchs, and M.~V. Chekhova.
\newblock Properties of bright squeezed vacuum at increasing brightness.
\newblock {\em Phys. Rev. Research}, 2:013371, 2020.

\bibitem{Spasibko:12}
K.~Yu. Spasibko, T.~Sh. Iskhakov, and M.~V. Chekhova.
\newblock Spectral properties of high-gain parametric down-conversion.
\newblock {\em Opt. Express}, 20(7):7507--7515, Mar 2012.

\bibitem{Yurke:86}
Bernard Yurke, Samuel~L. McCall, and John~R. Klauder.
\newblock {SU(2) and SU(1,1) interferometers}.
\newblock {\em Phys. Rev. A}, 33:4033--4054, Jun 1986.

\bibitem{Zakharov:18}
R.~V. Zakharov and O.~V. Tikhonova.
\newblock Photon spatial properties and correlations in nonclassical squeezed
  states of light carrying the orbital moment.
\newblock {\em Bulletin of the Russian Academy of Sciences: Physics},
  82(11):1388--1393, 2018.

\end{thebibliography}
\end{document}